\documentclass[
    ,final            
  ]
  {aipproc}

\layoutstyle{6x9}

\begin{document}

\title{Dynamical triggering of starbursts}

\author{F. Combes}{
  address={LERMA, Observatoire de Paris, 61 Av. de l'Observatoire, F-75014, Paris, France}
}

\begin{abstract}
Galaxy interactions/mergers, gravitational instabilities
and density waves, such as bars, are frequently invoked to
trigger starbursts. These mechanisms have been explored
through numerical simulations, with the help of various
star formation recipes.  Gravitational instabilities are 
necessary to initiate star formation, but the main 
trigger might be the gas flows, to provide sufficient
fuel in a short time-scale. Gas accretion is also 
acting on the dynamics, in favoring bars/spirals,
which will drive the gas inwards. Large amounts of
external gas accretion are required to explain the 
bar frequency, and this accretion rate can be provided by
the cosmic filaments, as supported
by cosmological simulations. Subsequent interactions 
can then trigger starbursts by driving
this accreted gas inwards.
\end{abstract}

\maketitle


\section{Observational evidence of dynamical triggering}

It is now widely accepted that strong galaxy interactions
can produce violent starbursts, and ultra-luminous 
infra-red galaxies (ULIRGs) are all mergers 
(e.g. Sanders \& Mirabel 1996).
 But interacting galaxies in general do not show intense starbursts
(Bergvall et al 2003), or only in their centers. The tidal parameter,
depending on mass and distance of companions, is not correlated
to any star formation tracer. Starbursts are rare in the local
universe, and much rarer than galaxy interactions. The only
perceptible effect is that interacting galaxies have more mass -
gas, dust \& young stars, and are more
concentrated to the center (Bergvall et al 2003).
One of the main conclusions that can be drawn for giant
galaxy starbursts is that galaxy
interactions is a necessary condition to
trigger them, but not sufficient. Certainly, other
parameters have to be taken into account, such as the 
gas content, the geometry/velocity of the encounter, 
and also the phases of the interaction, the starburst 
occuring only during about 10\% of the interaction
time, according to simulations (e.g. Mihos
\& Hernquist 1996).

For dwarf galaxies, interactions are even
not a necessary condition for triggering: starbursting
dwarfs have no more companions
(Brosch et al 2004), although it might
be different for Blue Compact Dwarfs (BCD, Hunter \& Elmegreen 2004).
Dwarf galaxies with starbursts present asymmetries,
and a possible trigger could just come from gas accretion:
the gas infalling in dark-matter haloes may experience  
sloshing and oscillations favoring compression and
instabilities. This mechanism might be 
relevant for primordial galaxies (Brosch et al 2004). 

\begin{figure}
  \includegraphics[width=.6\textwidth]{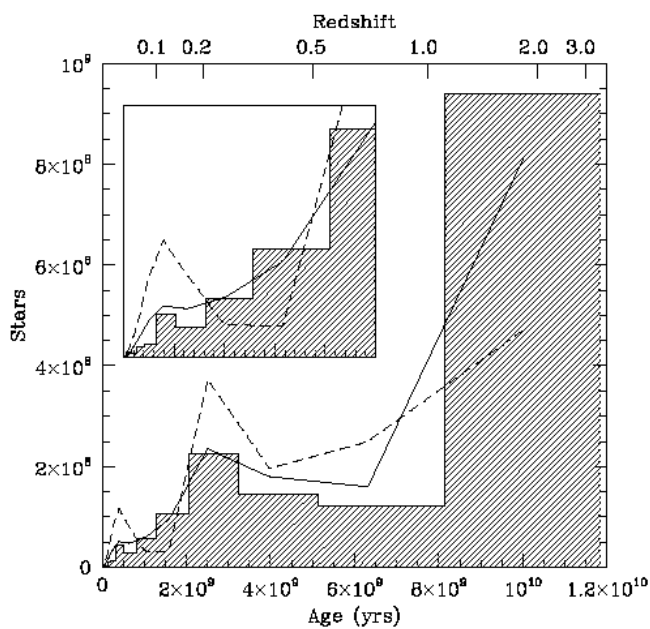}
  \caption{Star formation history in the SMC derived from 
a tidally triggered model (dash-line), or
a tidally triggered model with infall (solid line),
compared to the observed number
of stars at each age, from Zaritsky \& Harris (2004). The insert
is a zoom of the recent 2.5 Gyr.}
\label{zar-f11}
\end{figure}

One of the best evidence of triggering
can be obtained by tracing the fossil records
of the star formation history in nearby galaxies.
In the Small Magellanic Cloud, the number of stars
at a given age has been determined by
Zaritsky \& Harris (2004). They noticed that 
the derived SMC star formation history 
reveals some bursts corresponding to pericenters
of its orbit around the Milky Way. A model taking
the tidal triggering into account fits better 
the observations, although the quantitative influence
of the tide is uncertain, between 10 and 70\% of
the star formation could be tidal in origin,
cf figure \ref{zar-f11}. One of the problem in the fit
comes from the too large increase in star formation
about 3-4 Gyrs ago, which can only be explained by
external gas accretion.  A significant gas
infall, by about 50\% of the total gas mass, 
improves the fit, and accounts also for the 
observed age-metallicity relation.
The metallicity is not monotonously increasing
with decreasing age, but remains constant for old stars
until 3 Gyrs, indicating that dilution is required
from deficient gas infall.

\section{Dynamical processes}

The star formation rate in spiral galaxies is observed
to be proportional to the power n=1.5 of the gas density,
corresponding empirically to a global 
Schmidt law over the galaxy (Kennicutt 1998), although
the local Schmidt law is not observed. The SFR law is
the same for interacting and non-interacting galaxies,
whatever their triggers are.
This law has been interpreted through several
processes: Jeans instability can lead to this
exponent, star formation being proportional to
the gas density $\rho$ over the free-fall time scale,
scaling in $\rho^{1/2}$, resulting in the power 3/2, but 
also cloud-cloud collisions (Elmegreen 1998), or
star-forming contagion with feedback, leading to
chaotic conditions.

When strong starbursts are observed, they are in general
concentrated in the central regions, and large amounts
of gas must be fueled towards the center in a short enough
time to beat the feedback. The radial gas flows can be 
due to bars, or spiral density waves. 
Molecular gas concentrations are observed in
barred galaxies (e.g. Sakamoto et al 1999), and 
circumnuclear starbursts are frequently found in
resonant rings (e.g. Buta \& Combes 1996).

\vspace{-0.8cm}
\subsection{Star formation efficiency}
\vspace{-0.2cm}

The star formation efficiency (or SFE)
can be defined by the fraction of gas consumed in 
star formation per dynamical time.
In normal galaxies, the SFE is about 0.1-3\%, but in some
interacting galaxies and ULIRGs, it can be as high as 50\%.
The star formation rates in spiral galaxies range from
0.1 up to 1000 M$_\odot$/yr, depending on mass, but
SF can be about 100 times more efficient, in
triggered galaxies.

 The far infrared emission from dust L(FIR) is the best 
tracer of SF in enshrouded starbursts, and the gas
content is traced by the CO emission. The dense gas tracers
(CS, HCN) appear to vary linearly with L(FIR), and 
can be used also as SF tracer (Solomon et al 1992, Gao \& Solomon 2004).
While the SFE (FIR to CO ratio) may vary by a factor 100,
the HCN/CO emission ratio varies also in the same 
proportion, meaning that in average the gas is denser
in starbursts. The HCN/FIR ratio corresponds to star
forming cores in the Milky Way, and the ULIRGs can be
modelled with a large filling factor of SF cores. 
May be the necessary step to trigger a starburst 
is to form the dense cores, and
then the efficiency per unit core mass is universal
(Gao \& Solomon 2004, Fritze v. Alvensleben 2004).

\vspace{-0.8cm}
\subsection{Other dynamical processes}
\vspace{-0.2cm}

When two galaxies interact and their disks overlap,
there could be violent
collisions between clouds at velocities of the 
order of 300km/s; diffuse HI clouds collide, but
molecular GMCs not, due to their small filling factor.
However, they are compressed
by the created hot high pressure medium (Jog \& Solomon 1992).
Also, while GMCs are radially moving towards the center, 
they are compressed, provided that the ICM pressure increases
with decreasing radius (Jog \& Das 1996).
From stability arguments, if the Toomre Q-parameter
maintains constant over the whole disk, the
velocity dispersion and therefore the gas 
turbulence is higher in the center: the critical
velocity dispersion V$_{crit}$ (=3.36 G $\Sigma /\kappa$)
decreases with radius for constant rotation curves.
 
Similarly, in clusters, star formation could be
induced by the compression of GMCs by the 
intra-cluster hot gas (Bekki \& Couch 2003). This
might explain the starbursts occuring in
the spiral galaxies infalling into the cluster
Abell 1367 (Gavazzi et al 2003).

\section{Star formation Recipes}

Numerical simulations of galaxies use recipes
for star formation, as all sub-grid physics
(Katz 1992; Mihos \& Hernquist 1994). These
can be based on the Schmidt law, or the Jeans
instability, with some criteria taken into account,
like density thresholds, gas temperature,
convergent gas flows. These recipes can simulate
the dynamical triggering; simulated
galaxy interactions produce strong non-axisymmetry and
torques that drive the gas towards the center.
The strength of these perturbations
 depends essentially on the stability of the disk prior the
interaction, therefore on the bulge-to-disk ratio, while
the role of the orbital geometry appears secondary, provided
the interaction ends in a merger (Mihos \& Hernquist 1996).
However, many parameters involving the gas physics or
the stellar feedback can influence significantly the
SFR, while they are still uncertain.

\vspace{-0.8cm}
\subsection{Influence of gas stability}
\vspace{-0.2cm}

\begin{figure}
\begin{tabular}{cc}
   \begin{minipage}{6cm}
      \centering \includegraphics[width=5cm]{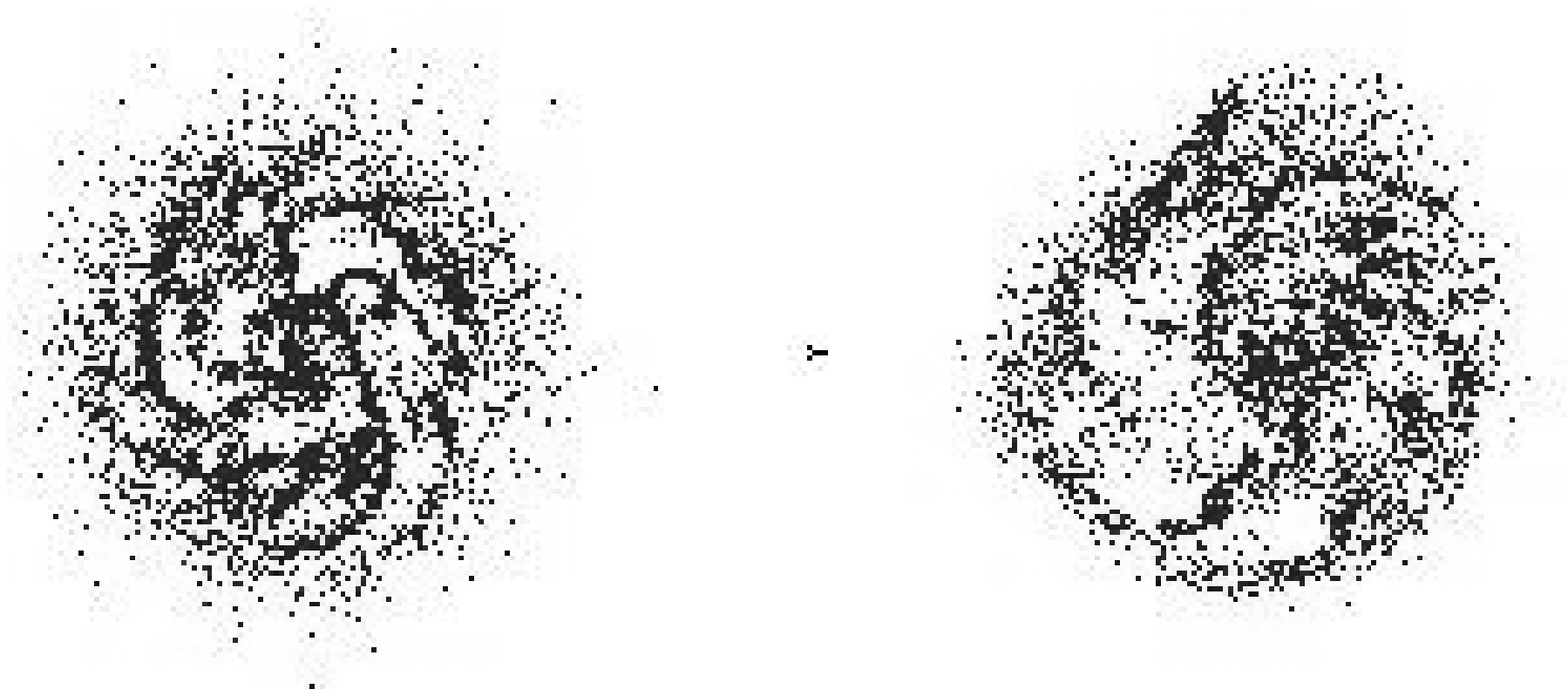}
   \end{minipage}
&
   \begin{minipage}{6cm}
      \centering \includegraphics[angle=-90,width=5cm]{combes-bh-f2b.ps}
   \end{minipage}
\\
   \begin{minipage}{6cm}
      \centering \includegraphics[width=5cm]{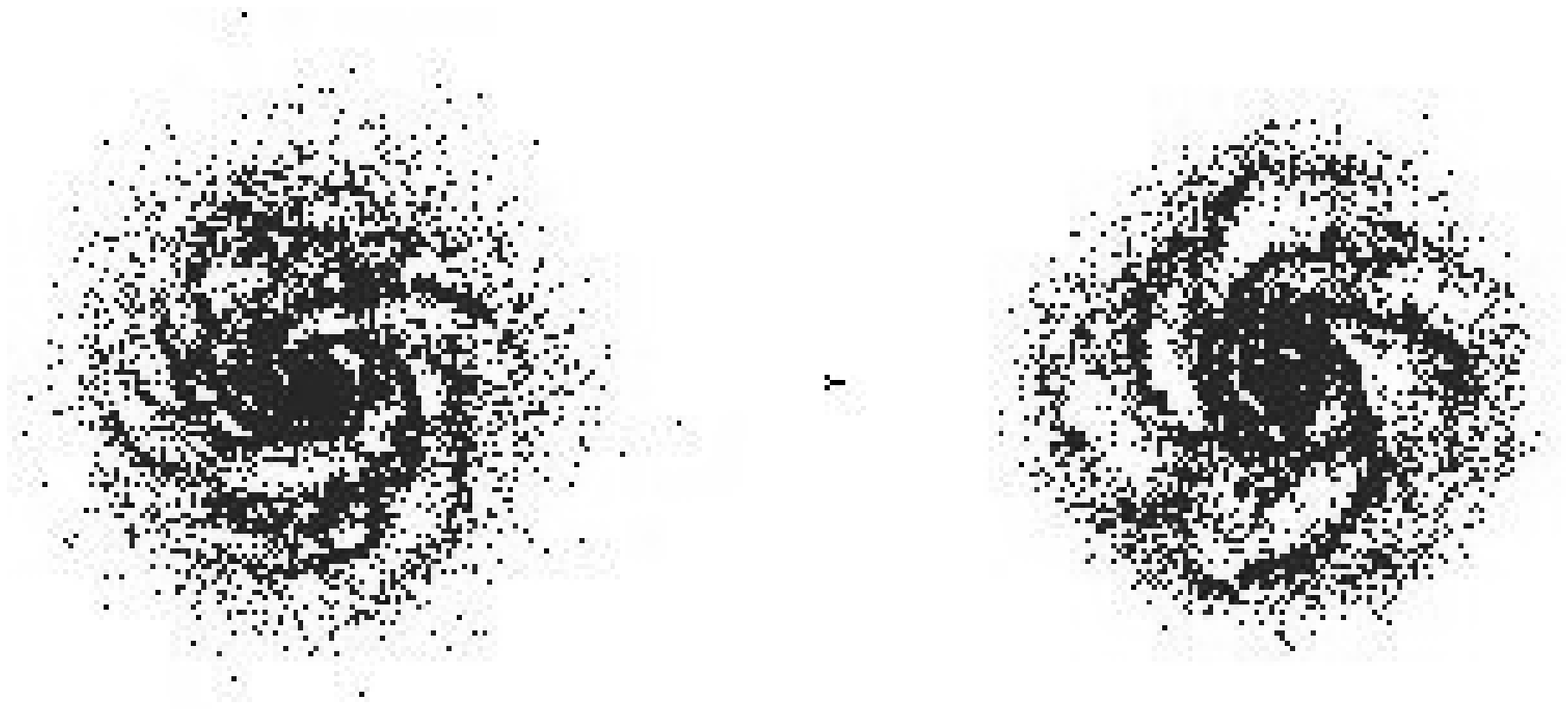}
   \end{minipage}
&
   \begin{minipage}{6cm}
      \centering \includegraphics[angle=-90,width=5cm]{combes-bh-f2d.ps}
   \end{minipage}
\\
\end{tabular}
      \caption{Comparison of two simulations of isolated
galaxies, with different Q-parameter for the gas (different
initial velocity dispersions). {\bf Top} Left: gas morphology
during the two first epochs of the simulation with Q$_g$ =0.3, separated by 200 Myr;
Right: corresponding SFR (fraction of the gas transformed into stars
per Myr) as a function of time.
 {\bf Bottom} Left: same for the simulation with Q$_g$ =0.8;
Rigth: Ratio of the SFR in these two simulations,
as a function of time. At the beginning,
the star formation rate is 3 times higher, for Q$_g$ =0.3.}
      \label{psfr}
\end{figure}

The small scale structure of the gas, depending on
its surface density in the galaxy, and its velocity 
dispersion, has a large influence on the SFR, modifying the 
fraction of dense clumps. Figure \ref{psfr} illustrates
how a small variation of initial Q$_g$ of the gas
can multiply the SFR by a factor 3.

The influence of the gas velocity dispersion can also
be included, to represent the shock-induced star formation
in interacting galaxies, since cloud-cloud collisions might
be involved in the mechanism. A modified Schmidt law 
in $\rho^n \sigma^{\beta}$ to take
into account the velocity dispersion $\sigma$ has been introduced
by Brosche \& Lentes (1985). It was applied by
Barnes (2004) to the Mice simulations, and revealed a
better match of the observations, with the exponent
 $\beta$ = 0.5.

\vspace{-0.8cm}
\subsection{Influence of multi-phase gas, feedback}
\vspace{-0.2cm}

Hydro-SPH simulations ignore the various phases, and smooth the density 
contrast: hot diffuse gas is prevented to exist together with dense cold
clumps; the dense gas evaporates, while the hot gas cools
exaggerately.  Marri \& White (2003)
attempt to separate two gas phases in the
SPH code (with neighbor particles searched separately)
taking into account mass exchange between the two phases:
this of course changes significantly the star formation 
history of a spiral galaxy.
The star formation should also be limited by stellar feedback,
which is phenomenologically taken into account through
mass loss and energy reinjected locally around star-forming regions.
However the large-scale fountain effect, or galactic winds, 
and large-scale outflows are not simulated correctly
(Thacker \& Couchman 2001).

\section{Importance of gas infall}

Many observations point towards an almost
stationnary star formation history in spiral galaxies
for intermediate Hubble types
(Kennicutt et al 1994, Brinchmann et al 2004).
However, even taking into account the stellar mass loss, an isolated 
galaxy should have an exponentially decreasing star
formation rate. To fuel the star formation,
galaxies depend on external accretion. Mergers with
gas rich dwarf companions are not sufficient;
for instance all objects falling
now on the Milky Way represent less than
1/400 of the mass of the Galaxy (Ibata et al 2001).
Galaxies must accrete large amounts of gas mass
along their lives (Brinchman et al 2004).

The analysis of stellar populations in a large sample of galaxies
(SDSS, Heavens et al 2004) have shown that
massive galaxies have formed most of their stars at early times,
while dwarf galaxies are still forming now.
Only galaxies of intermediate masses could have in average maintained
their star formation rate over a Hubble time.

\vspace{-0.8cm}
\subsection{Simulations in a cosmological context}
\vspace{-0.2cm}

When simulated inside
the filamentary cosmic web of gas and dark matter,
the star formation rate of galaxies 
is no longer exponential. The star formation is both
fueled by galaxy mergers, and in the mean time,  external
gas accretion (e.g. Tissera 2000, Nagamine et al 2004).
One of the essential role of companions is to trigger 
star formation, by driving this accreted gas inwards. 
Semi-analytical simulations, constrained by the luminosity
function of galaxies and the Tully-Fisher relation,
also derive an almost constant SFH for middle types 
(Somerville \& Primack 1999).

\vspace{-0.8cm}
\subsection{Constraints from bars and spirals}
\vspace{-0.2cm}

Recently, the bar frequency in nearby galaxies
have been quantified from near-infrared surveys
(e.g. Block et al. 2002), and is much higher than
expected, given the destruction of bars in spiral
galaxies with interstellar gas (Bournaud \& Combes 2002).
To explain the observations, bars have to be reformed, and
this is possible only through continuous gas accretion at a large rate.
The observed bar frequency, combined with the
prediction of dynamical models, allows to quantify the accretion rate
 (Block et al. 2002). An intermediate type spiral galaxy must
double its mass in about 10 Gyr.
Cold gas accretion is required to increase the 
self-gravity of the disk, and decrease the bulge-to-disk ratio.
Dwarf companions accretion cannot play this role, since they
heat the disks. Also major mergers develop the spheroids.
Instead the filaments in the near environment of galaxies
are the source of continuous cold gas accretion.
 The accretion rate derived in cosmological 
simulations is compatible with the mass doubling in 10 Gyr.
 Cosmological accretion can explain bar reformation.

\section{Conclusions}

Star formation depends essentially on the gas supply
and is only slightly influenced by the large-scale dynamics.
Galaxy interactions act to drive the accreted gas radially inwards
 and trigger central starbursts. The
accretion of gas not only regulates the star forming history in galaxies 
but also their dynamical state  (bars, spirals, warps, m=1...), and both
phenomema interact in a feedback loop.
For galaxies in the field, accretion is dominant, and explain bars and spirals, and
the constant star formation rate for intermediate types.
In rich environments, galaxies experience quicker evolution, 
mergers are much more important, and
secular evolution of galaxies through gas accretion
is halted at z$\sim 1$, since galaxies
are stripped from their gas reservoirs.



\end{document}